\documentclass[namedreferences]{solarphysics}

\usepackage[hyperref,optionalrh,showbiblabels]{spr-sola-addons} 
\usepackage{graphicx}        
\usepackage{color}           
\usepackage{breakurl}        

\newcommand{\aap}{    {\it Astron. Astrophys.}}

\newcommand{\apj}{    {\it Astrophys. J.}}

\newcommand{\solphys}{{\it Solar Phys.}}
 
\newcommand{\ssr}{    {\it Space Sci. Rev.}} 
\chardef\us=`\_

\begin{document}

\begin{article}
\begin{opening}

\title{Magnetic Power Spectra of Emerging Active Regions}

\author[email={olzk@mail.ru}]{\inits{O.K.}\fnm{Olga K.}~\lnm{Kutsenko}\orcid{0000-0002-6670-7119}}
\author[corref, email={alex.s.kutsenko@gmail.com}]{\inits{A.S}\fnm{Alexander S.}~\lnm{Kutsenko} \orcid{0000-0002-1196-5049}}
\author[email={vabramenko@gmail.com}]{\inits{V.I}\fnm{Valentina I.}~\lnm{Abramenko}\orcid{0000-0001-6466-4226}}

\address{Crimean Astrophysical Observatory of Russian Academy of Sciences, p/o Nauchny, 298409, Crimea, Russia}

\runningauthor{O.K~Kutsenko \textit{et al.}}
\runningtitle{Magnetic Power Spectra of Emerging ARs}

\begin{abstract}
Magnetic field data provided by the \textit{Helioseismic and Magnetic Imager} on board the \textit{Solar Dynamics Observatory} were utilized to explore the changes in the magnetic energy of four active regions (ARs) during their emergence. We found that at the very early stage of the magnetic flux emergence, an abrupt steepening of the magnetic power spectrum takes place leading to rapid increase of the absolute value of the negative spectra power index $\alpha$ in $E(k)$ $\sim$ $k^{\alpha}$. As the emergence proceeds, the energy increases at all scales simultaneously implying that elements of all sizes do appear in the photosphere. Meanwhile, the energy gain at scales larger than $\approx$10 Mm prevails as compared to that at smaller scales. Both direct (\textit{i.e.}, fragmentation of large structures into smaller ones) and inverse (\textit{i.e.}, merging of small magnetic features into larger elements) cascades are readily observed during the emergence. However, in the case of inverse cascade, the total energy gained at large scales exceeds the energy loss at smaller scales assuming simultaneous appearance of large-scale magnetic entities from beneath the photosphere. We conclude that most of the time the energy may grow at all scales. We also cannot support the point of view regarding the dominant role of the inverse cascade in the formation of an AR. Although the coalescence of small magnetic elements into larger pores and sunspots is observed, our analysis shows that the prevailed energy contribution to an AR comes from emergence of large-scale structures.

\end{abstract}
\keywords{Active Regions, Magnetic Fields; Turbulence}
\end{opening}

\section{Introduction}
     \label{S-Introduction} 

Last decades a considerable attention was paid to both theoretical and observational study of magnetic flux emergence on the Sun. The observational picture of emergence can be described as follows. First visible imprints of an emerging active region (AR) are small magnetic features of different polarities that appear at the surface and move apart from the emergence site. A strong transverse magnetic field of hundreds or thousands of gauss is observed between the features \citep{Brants1985, Zwaan1985, Lites1998}. Magnetic elements of opposite polarities migrate to different sides of the emergence site and coalesce to form pores. When the future footpoints of an AR become separated far enough, small magnetic dipoles start to appear throughout the emergence site and move toward the footpoints  \citep{Centeno2012}. 
These magnetic dipoles, labeled by \cite{Bernasconi2002} as magnetic dipolar features (MDF), are presumably series of undulatory $\Omega$- and $U$-loops of almost horizontal magnetic flux ropes that connect the AR footpoints. The ropes are seen as elongated darkenings in the photospheric continuum intensity  \citep[\textit{e.g.,}][]{Strous1999}. MDF are stitches where the magnetic flux is still tied to the upper photosphere \citep{Bernasconi2002}. Gradually the reconnections of elementary $U$-loops occur. The magnetic flux tubes lose excessive mass and the arc filament system rises up above the surface. Pores of the same polarity could merge to sunspots.
More detailed reviews on magnetic flux emergence and AR formation can be found elsewhere \citep[\textit{e.g.,}][]{vanDrielGesztelyi2015, Cheung2017}.

The above description of the AR formation might suggest that the magnetic flux emerges predominantly by small portions rather than as a monolithic  structure. Recent numerical simulations support this qualitative suggestion. For a review of theoretical aspects of flux emergence see, e.g., \cite{Fan2009, Cheung2014}. Here we mostly focus on recent results reported by \cite{Chen2017} since they performed the most realistic comprehensive numerical 3D magnetohydrodynamical simulations of magnetic flux emergence so far.

\cite{Chen2017} considered the emergence of magnetic flux bundle from the uppermost layers of the convection zone to the corona. Magnetic and flow fields, as extracted from a spherical global dynamo simulation by \cite{Fan2014}, were coupled to the lower boundary of their computational domain. The simulated magnetic field and continuum intensity images at the solar surface are quite similar to magnetograms and continuum intensity maps observed by high-resolution instruments. Authors further argue that a magnetic flux bundle rises as a coherent structure through the convection zone before reaching the uppermost several Mm below the photosphere. At these depths, the flux bundle fragments into magnetic tubes that emerge individually. During the first stages of magnetic flux emergence, only small-scale granular-size magnetic elements appear at the solar surface. They gradually coalesce to form larger magnetic flux concentrations that appear as pores in the intensity map. As the emergence proceeds, pores seem to merge to sunspots.

Both theoretical and observational results mentioned above suggest that coalescence of small-scale magnetic features with formation of larger structures is dominant during the magnetic flux emergence. This process is referred to as inverse turbulent cascade \citep[\textit{e.g.},][]{Biskamp1993}, when the energy from small scales transfers to larger scales. The preliminary observational evidence of the inverse cascade was reported, for the first time, by \cite{Hewett2008} basing on low-resolution magnetograms of a peculiar, fast-emerging strong AR NOAA 10488. The authors used full-disk magnetograms obtained by the \textit{Michelson Doppler Imager} on board the \textit{Solar and Heliospheric Observatory} \citep[SOHO/MDI, ][]{Scherrer1995}. At the same time, since emergence occurs in a highly turbulent medium of solar plasma, an opposite process -- the direct cascade, \textit{i.e.}, fragmentation of large magnetic structures into smaller ones -- should take place as well. Using modern magnetographic data for a set of ARs, we aim to explore the interplay between magnetic energy loss and gain at different spatial scales and to find out whether the inverse cascade is a common property of emergence.

The magnetic energy spectrum is a very useful technique to estimate the amount of magnetic energy stored at different spatial scales, as well as the energy transfer along scales. In this study, we use a method for the 1D power spectrum acquisition \citep{Abramenko2001, Abramenko2005}. The method is based on the Fast Fourier Transform of 2D-magnetograms.

Uninterruptible high-cadence and high-spatial-resolution magnetic field data provided by the \textit{Helioseismic and Magnetic Imager} on board the \textit{Solar Dynamics Observatory} \citep[SDO/HMI,][]{Scherrer2012, Schou2012} allowed us to explore in details the changes in the magnetic energy spectrum during the AR emergence. We focus on time changes of the power index (the slope of the spectrum), as well on signatures of energy decrease at large scales with simultaneous increase at smaller scales, \textit{i.e}, direct cascade, and on opposite process, \textit{i.e.}, inverse cascade.

\section{Data Reduction and Power Spectra Calculation}
\label{S-methods}

We employed line-of-sight (LOS) magnetic field measurements provided by SDO/HMI. The instrument is a full-disk filtergraph that routinely acquires narrow-band filtergrams of the full disk of the Sun by two CCDs (``front'' and ``side'' cameras, \citealp{Schou2012, Liu2012}). The spatial resolution of 4096$\times$4096 pixels filtergrams is 1'' with a 0.5'' pixel size. The side camera is aimed at vector magnetic field measurements. It performs filtergrams acquisition in six polarization states at six wavelengths positions within the photospheric spectral line Fe {\sc i} 6173 \AA \citep{Norton2006}. It takes 135 s to acquire a full set of 36 filtergrams. In order to increase the signal-to-noise ratio, the Stokes parameters are computed every 720 s from filtegrams averaged over 1350 s of data \citep{Couvidat2012b}.

LOS observables are produced from filtergrams acquired in right-circular (RCP) and left-circular polarization (LCP) states (12 of 36 filtergrams). To derive full-disk LOS magnetograms used in this work, the so-called MDI-like algorithm is applied \citep{Couvidat2012a}. Briefly, the magnetic field is proportional to the difference of Doppler velocities obtained from the observed spectral line in RCP and LCP. Look-up tables are used to correct the derived Doppler velocities for the finite width of the filter profiles, orbital velocity of the satellite, asymmetry of the spectral line, \textit{etc}. The filling factor in SDO/HMI magnetic field measurements is assumed equal to unity. Consequently, the values in magnetogram pixels are magnetic flux densities over the pixel area rather than magnetic field strength. The noise in the full-disk 720-s LOS magnetograms varies between 4.8 Mx~cm\textsuperscript{-2} at the disk center and 7.9 Mx~cm\textsuperscript{-2} near the solar limb \citep{Liu2012}.

SDO/HMI LOS magnetic field become saturated every 12 hours if the magnetic flux density in a pixel exceeds 3200 Mx~cm\textsuperscript{-2} \citep{Liu2012}. The reason for this saturation is the spectral line shift beyond a well-calibrated part of the calibration curve due to combined action of large Zeeman and Doppler shifts. The latter is determined by the velocity of the geostationary satellite relative to the Sun. The 12- and 24-hour periodicities might also contribute to weaker LOS magnetic fields \citep{Kutsenko2016}. However, we did not reveal the existence of these periodicities in our analysis.

Full set of acquired Stokes parameters is also inverted \citep{Borrero2011} and disambiguated \citep{Metcalf1994, Leka2009} to retrieve the vector magnetic field. The data are available in the form of full-disk vector magnetograms or vector magnetic field patches of ARs \citep{Bobra2014, Hoeksema2014}. The noise in vector magnetic field data is about two orders of magnitude higher than that in LOS observables (values of magnetic flux density below 220 Mx~cm\textsuperscript{-2} in the vector magnetic field measurements should be assumed as noise according to \citealp{Bobra2014}). This is the reason that motivated us to use LOS magnetograms instead of vector magnetic field data in this work.

For each analyzed AR, we compiled a series of LOS magnetic field patches extracted from full-disk magnetograms.  Emerging ARs, as well as the quiet-Sun areas where they appeared, were tracked as long as the centers of the patches were located within $\pm$30 heliographic degrees from the central meridian. To further minimize the projection errors, the $\mu$-correction for each pixel in the patches was performed \citep[\textit{e.g.},][]{Leka2017} before calculation of power spectra. The total unsigned magnetic flux, $\Phi$, was also computed as a sum of absolute magnetic flux densities in pixels multiplied by the corrected area of a pixel. The summation was performed only over those pixels where the absolute magnetic flux density value exceeded 18 Mx~cm\textsuperscript{-2} -- the triple standard deviation noise level \citep{Liu2012}.

To calculate the magnetic power spectra of ARs, we use an approach developed in \cite{Abramenko2001, Abramenko2005}. Briefly, the 2D power spectrum of a LOS magnetogram acquired at time $t$ was calculated as a squared Fast Fourier Transform of the magnetogram. Then 1D power spectrum $E(k, t)$ was then retrieved. Here $k=2\pi/r$ is the wavenumber and $r$ is the spatial scale (in Megameters, Mm). $E(k, t)$ was calculated by summing the 2D-spectrum over a thin ring in the 2D-space of wavenumbers \citep[see][for details]{Abramenko2005}. Usually, the inertial range is well visible in each spectrum as a linear interval in a double-logarithmic plot, so that $E(k, t)\propto k^{\alpha}$ with the power index $\alpha(t)$. We computed $\alpha(t)$ from the best linear approximation to the data points inside a range $\Delta r$=2.7--10 Mm. The choice of the range is based on results reported in \citep{Abramenko2001, Abramenko2005}: for SOHO/MDI high resolution data, at scales lower than approximately 3 Mm the power index calculation is marginally affected by the influence of noise and insufficient resolution of the telescope. We found that in the case of SDO/HMI data, the lower limit can be extended to 2.7-2.4 Mm. We further use the value of 2.7 Mm as the most reliable high-frequency limit of the inertial range. \cite{Mandage2016} argued that the large-scale limit of the inertial range depends on the AR being analyzed and could be extended above 10 Mm. The extension of the inertial range is obviously caused by presence of large-scale magnetic structures such as sunspots. However, since we intend to obtain homogeneous estimates of the power index $\alpha$ for both quiet, pre-emergence photosphere and AR areas, we adopted the large-scale limit of the inertial range $r$=10 Mm where the influence of large sunspots on the spectrum is negligible.

To analyze time variations of the energy transfer along the spectrum, we calculated the partial time derivative of the power spectrum, $\partial E(k,t)/ \partial t$: for each $k$ the derivative along the time axis was computed. Five-point stencil was used to perform numerical differentiation. Using the customary notation in turbulence theory \citep[\textit{e.g.}, ][]{Biskamp1993}, we will refer to this function as nonlinear energy transfer function. 

\section{Results} 
      \label{S-results}      

Four emerging NOAA ARs under study are listed in Table~\ref{table1}. Second column shows the beginning and the end of analyzed time interval. The total unsigned magnetic flux at the maximum of AR development is presented in the last column. All ARs developed into moderate bipolar magnetic structures obeying Hale polarity law and Joy's law. Only one of them, the largest AR 11726, revealed a peculiarity: the following-polarity spot was a dominating feature between other sunspots in the group. The time interval of our analysis covered the pre-emergence phase as well when no signatures of emerging magnetic structures appeared at the photosphere.

\subsection{Evolution of the Power Index $\alpha(t)$}

\begin{figure}    
	\centerline{\includegraphics[width=1\textwidth,clip=]{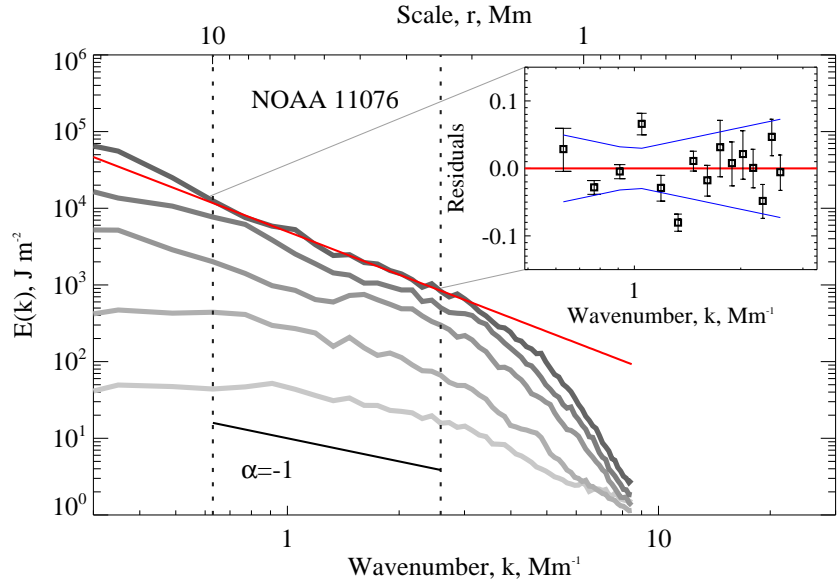}
	}
	\caption{
	 The evolution of the magnetic power spectra as AR NOAA 11076 emerges. The spectra are plotted with the cadence of 24 hours, from lightest to darkest grey curve. The first spectrum is derived from the magnetogram acquired at the start moment listed in Table~\ref{table1}. Dotted vertical lines enclose the wavenumber range where the power index $\alpha$ was determined by the least-square fitting of the spectrum by a linear function $y=ak+b$ in a double-log scale. For comparison, a segment with the slope of -1 is shown. The red line demonstrates the best-linear fitting of the last (darkest) spectrum curve. The residuals of the fit $\log_{10}(E(k)) - \log_{10}(y)$ are shown in the upper-right panel. Blue curves show 95\% confidence intervals of the fit. The error bars are estimated as maximum absolute variations of the spectrum points within the time interval of one hour centered at the time when the darkest spectrum was acquired.
	}
	\label{fig1}
\end{figure}

\begin{figure}    
	\centerline{\includegraphics[width=1\textwidth,clip=]{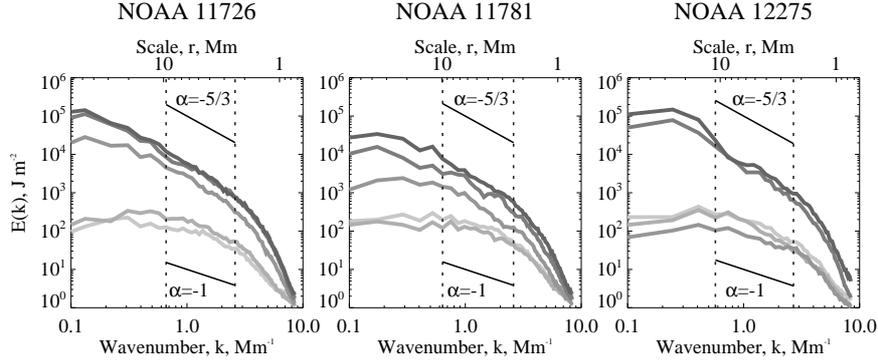}
	}
	\caption{
		The same as in Figure~\ref{fig1} for AR NOAA 11726 (left), 11781 (middle), and 12275 (right). The first spectrum on each panel is derived from the magnetogram acquired at the start moment listed in Table~\ref{table1}. Dotted vertical lines enclose the wavenumber range where the power index $\alpha$ was determined by the best-linear fitting of the curve. For comparison, segments with the slopes of -1 and -5/3 are shown.
	}
	\label{fig2}
\end{figure}

Figures~\ref{fig1} and \ref{fig2} demonstrate how the magnetic power spectrum evolves as the magnetic structure emerges. The spectra are separated by the 24-hour interval starting from the lightest grey curve in each panel. As an AR emerges, the power spectrum becomes steeper, and positive energy increment over a day is visible at all scales. The linear range of 2.7--10 Mm, where the power index $\alpha$ was calculated, is enclosed between dotted vertical segments. During the pre-emergence phase, all ARs demonstrate the spectrum with  $\alpha$ close to -1, which is expected for the magnetic power spectrum in the quiet-sun photosphere \citep{Abramenko2001}. As the AR emerges, the spectrum becomes steeper and value of $\alpha$ reaches the Kolmogorov's -5/3 and even more \citep[see, \textit{e.g.},][]{Abramenko2005}.

\begin{figure}    
	\centerline{\includegraphics[width=1\textwidth,clip=]{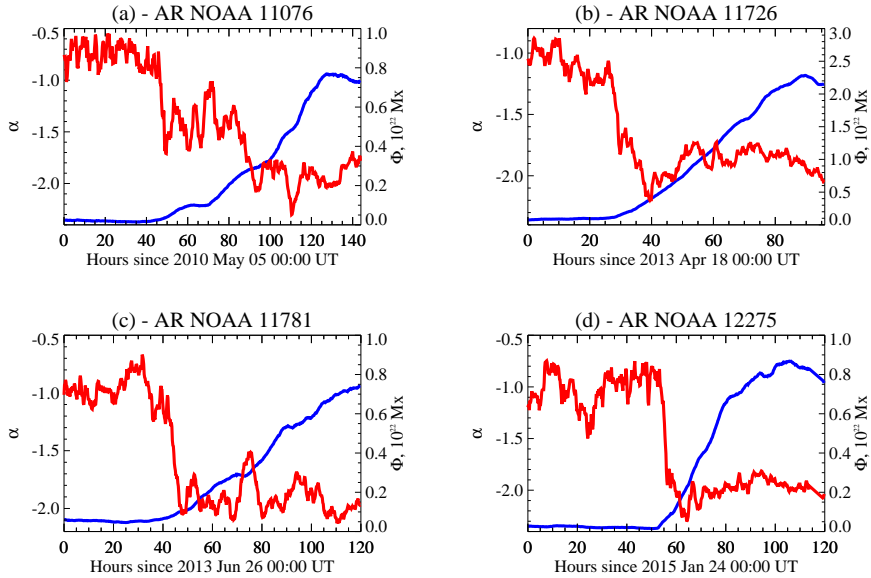}
	}
	\caption{
		Time changes of the total unsigned magnetic flux (blue curve) and power index, $\alpha$, (red curve) for emerging ARs.
	}
	\label{fig3}
\end{figure}

Time variations of the power index $\alpha$, along with the total unsigned flux, are shown in Figure~\ref{fig3}. The pre-emergence magnitude of $\alpha \approx -1$  abruptly changes to lower magnitudes (higher absolute values) at the time of the very beginning of the flux growth. As the emergence progresses, the magnitude of $\alpha$ does not change much undulating around the level of -2. We emphasize that at the time of rapid drop of $\alpha$ only the first small magnetic features of the future AR are visible on the photosphere. Such a drop can serve as an indicator of the near-future appearance of a (at least) moderate  active region.

\begin{table}
	\caption{List of ARs under study and some of their parameters.}
	\label{table1}
	\begin{tabular}{ccc}     
		\hline                   
		NOAA AR & Dates of & Peak total unsigned\\
		& observations & flux, $10^{22}$ Mx \\
		\hline
		11076 & 2010, May 29/00:00 UT -- Jun 03/23:48 UT & 0.8  \\
		11726 & 2013, Apr 18/00:00 UT -- Apr 21/23:48 UT & 2.3  \\
		11781 & 2013, Jun 26/00:00 UT -- Jun 30/23:48 UT & 0.7  \\
		12275 & 2015, Jan 24/00:00 UT -- Jan 28/23:48 UT & 0.9  \\
		\hline
	\end{tabular}
\end{table}

\subsection{Evolution of the Nonlinear Transfer Function $\partial E(k,t)/ \partial t$ }

\begin{figure}    
	\centerline{\includegraphics[width=1\textwidth,clip=]{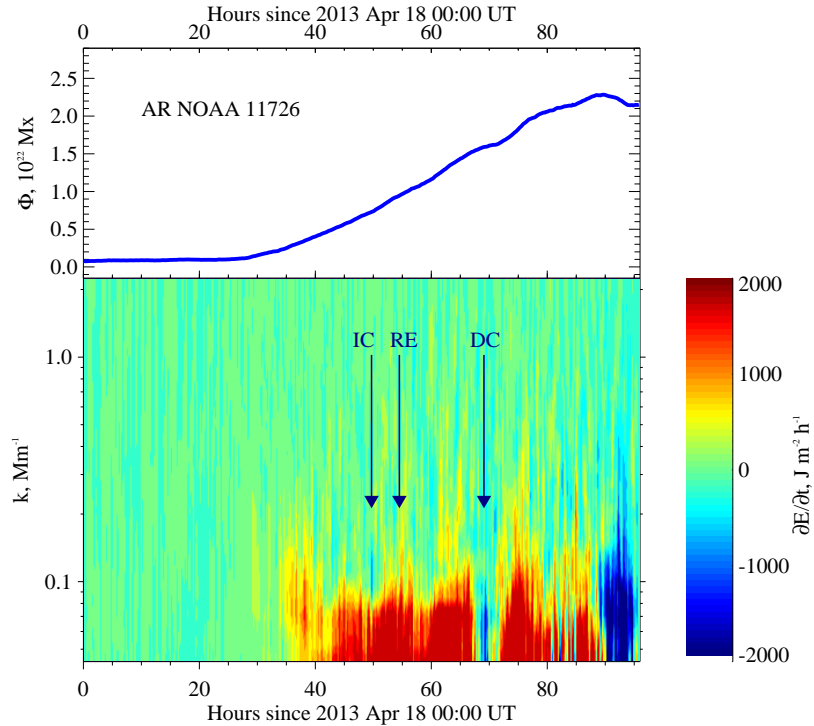}
	}
	\caption{
		The $k$--$t$ plot made along the nonlinear transfer function $\partial E(k,t)/ \partial t$ for AR NOAA 11763. The energy enhancement (positive derivative) corresponds to shades of red, and the energy losses (negative derivative) are indicated with blue. The range of scaling is shown in the color bar.
		Arrows indicate the instants of well pronounced direct cascade (DC), inverse cascade (IC), and an example of regular emergence (RE). For comparison, the total unsigned flux growth along the same time line is presented on the top.   		
	}
	\label{fig4}
\end{figure}

\begin{figure}    
	\centerline{\includegraphics[width=1\textwidth,clip=]{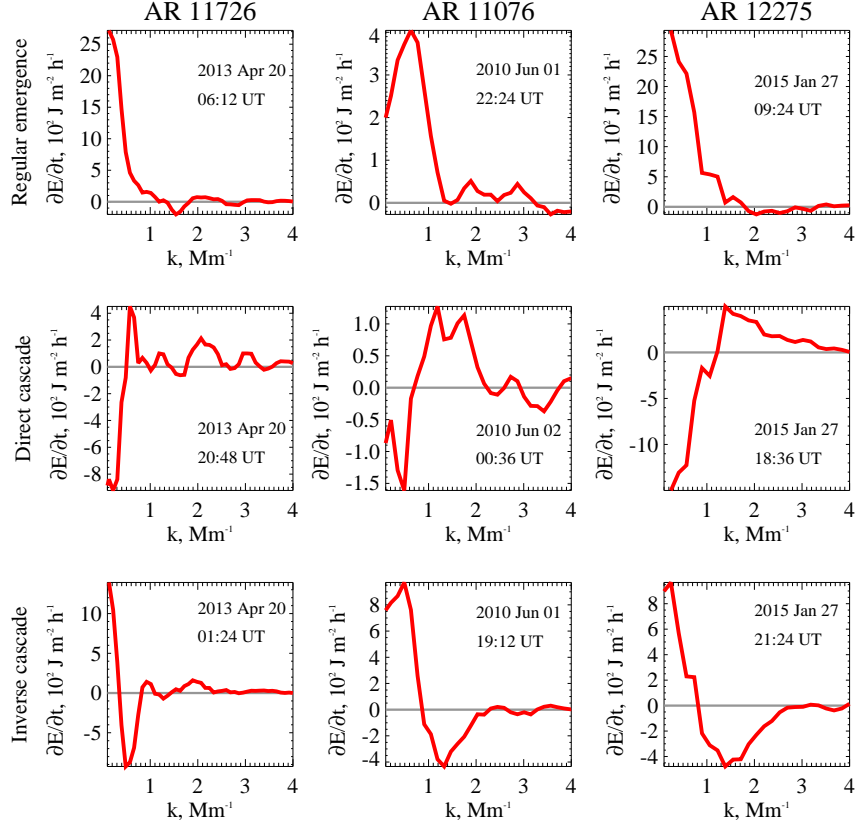}
	}
	\caption{
		The slices of the $\partial E(k,t)/ \partial t$ function at certain moments illustrating three regimes (regular emergence (upper row), direct cascade (middle row), and inverse cascade (bottom row)) for ARs NOAA 11726 (left column), 11076 (middle column), and 12275 (right column). The slices of the $\partial E(k,t)/ \partial t$ function for AR NOAA 11726 are shown at the moments indicated by blue arrows in Figure~\ref{fig3}.
	}
	\label{fig5}
\end{figure}

Since the nonlinear energy transfer function $\partial E(k,t)/ \partial t$ is a time derivative of the magnetic energy spectrum, it allows us to better reveal how energy changes at different scales or wavenumbers $k$ from one consecutive magnetogram to another. Negative values of the nonlinear energy transfer function at certain scales imply energy loss while positive values are observed if the energy gain took place. The visual representation of $\partial E(k,t)/ \partial t$ allows one to see the temporal interplay (say, simultaneous energy increase at larger scales and energy decrease at smaller scales) of energy cascades between different spatial scales. An example for NOAA AR 11763 is shown in Figure~\ref{fig4}. Shades of red show the energy growth at a given scale on a given time, whereas blue stripes indicate the energy losses. Some individual $\partial E(k,t)/ \partial t$ functions are presented in Figure~\ref{fig5}. When we observe the energy gain at small scales with simultaneous energy loss at large scales, this is a moment of well pronounced direct cascade, when energy  might be transferred from large to smaller scales. In other words, large-scale entities might be fragmented. Examples of this situation are visible in Figure~\ref{fig4} (a yellow strip on small scales is observed simultaneously with blue stripes at large scales; one of them is marked by DC-arrow), and in Figure~\ref{fig5} (middle row). The opposite process -- inverse cascade -- is illustrated in Figure~\ref{fig4} by the IC-arrow and in Figure~\ref{fig5} (bottom row). On the whole, blue and red-yellow stripes in the $k$--$t$ plot constantly alternate each other along the time axis, which means very inhomogeneous regime of emergence: energy gain can be replaced by energy loss (and \textit{vice versa}).  
   
Short intervals when inverse and direct cascades are dominant and well visible on the $k$--$t$ plot are rather rare. The regular emergence (RE) regime (Figure~\ref{fig4}, RE-arrow) is the energy gain along the entire spectrum, especially at scales above 3 Mm (approximately $k<$ 2, Figure~\ref{fig5}, top row). Of cause, both cascades are continuously present in turbulent medium, and they are competing for the spectrum formation. Anyway, we do not observe any prevalence of the inverse cascade (formation of large magnetic concentrations predominantly due to convergence of small ones) during the emergence phase. As for the right-end boundary of about 3 Mm, this might be related to the influence of noise and insufficient resolution on the power spectrum calculations \citep{Abramenko2001}.
 
Cadence of analyzed data is 12 minutes, so that consecutive spectra in Figure~\ref{fig4} are separated by the 12-minute intervals. This interval might be not sufficient to properly reveal reorganizations accompanying the dominant turbulent cascade action. We thus performed the following experiment to visualize the existence of the inverse cascade on longer time interval. 

\begin{figure}    
	\centerline{\includegraphics[width=1\textwidth,clip=]{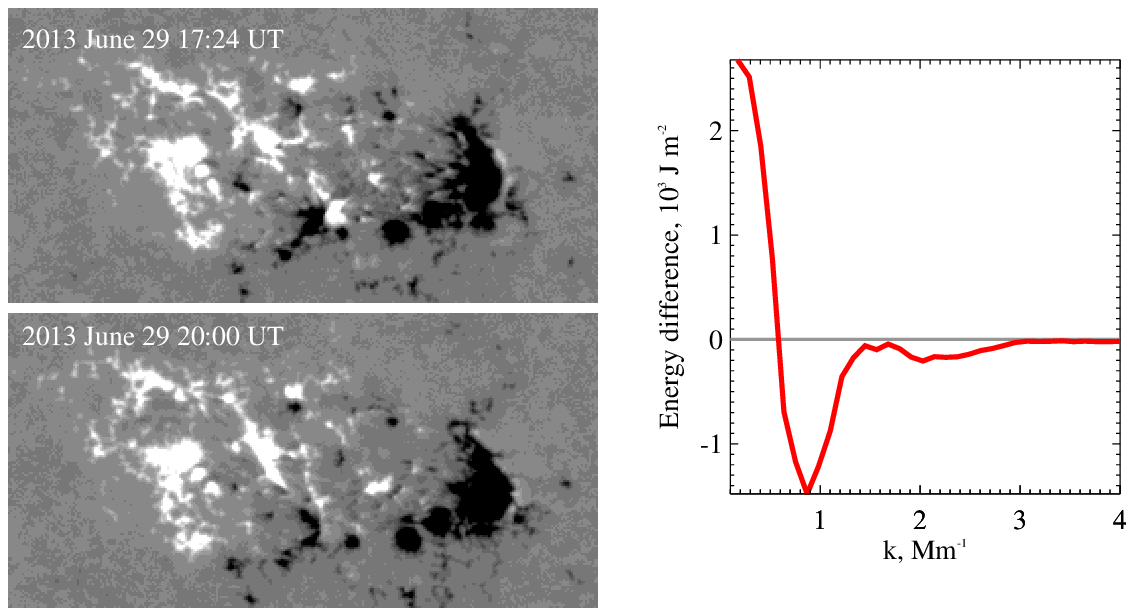}
	}
	\caption{
		Illustration of the inverse cascade from 2.5-hour observations. Left -- SDO/HMI LOS magnetograms of AR NOAA 11781, acquired on 2013 June 29 at 17:24 UT ($t_0$, top) and at 20:00 UT ($t_1$, bottom). The size of the field-of-view is 150 by 75 arcsec. The magnetograms are scaled from -500~Mx~cm\textsuperscript{-2} (black) to 500~Mx~cm\textsuperscript{-2} (white). Right -- The total energy injection at different wavenumbers accumulated from $t_0$ to $t_1$ and calculated as $\int_{t_0}^{t_1} \partial E(k,t)/ \partial t \mathrm{d}t$.
	}
	\label{fig6}
\end{figure}

Figure~\ref{fig6} shows SDO/HMI LOS magnetograms of AR NOAA 11781 acquired on 2013 June 29. Small-scale elongated magnetic features are clearly seen on the magnetogram acquired at $t_0$ = 17:24 UT (top-left panel). By the time $t_1$ = 20:00 UT (bottom-left panel), these magnetic features merged with the leading negative polarity spot, \textit{i.e.} the inverse cascade is clearly distinguished by visual means. 
 
We calculated the total energy injection at different spatial scales during the aforementioned interval as $\int_{t_0}^{t_1} \partial E(k,t)/ \partial t \mathrm{d}t$. The resulting curve is shown in the right panel of Figure~\ref{fig6}. Indeed, the energy decrease at small scales, as well as co-temporary energy increase at large scales, is observed. However, the total energy injected at large scales (which can be evaluated as an area above the positive section of the curve in the right panel of Figure~\ref{fig6}) is almost twice  higher than that removed from the small scales. Obviously, along with the inverse cascade, simultaneous appearance of large-scale magnetic entities does take place.

\section{Conclusions and Discussion} 
      \label{S-conclusuions}      

We used SDO/HMI LOS magnetograms for four emerging ARs  to explore the inter-scale magnetic energy transfer on the basis of the magnetic power spectrum technique. Analyzing power spectra and their variations in time, we found the following:
\begin{itemize}

\item At the very early stage of the magnetic flux emergence, when only tiny magnetic features are visible in the photosphere, an abrupt steepening of the magnetic power spectrum occurs, which is well pronounced in increasing of the absolute value of the negative power index $\alpha$ in $E(k)$ $\sim$ $k^{\alpha}$. The result can be useful for early diagnostics of emerging active regions. 
 
\item Most of the time the energy increases at all scales simultaneously, implying that elements of all sizes do appear in the photosphere after the emergence onset. The energy gain at larger scales prevails as compared to that at smaller scales.
 
\item Evidences of the direct cascade (when energy from large scales can be transferred to small scales, \textit{i.e.}, fragmentation), as well as signatures of the opposite inverse cascade, were revealed. However, in the case of the inverse cascade, the total energy gained at large scales exceeds the energy loss at smaller scales, so that along with coalescence of small features, simultaneous appearance of large-scale magnetic entities does take place.

\item The process of emergence displays highly intermittent in time nature: short periods of energy gain in some scale interval can be quickly replaced by energy losses, and \textit{vice versa}, \textit{i.e.}, the trade interactions between scales are always going on.
 
\end{itemize}
 
We thus observe that during the emergence the magnetic energy grows at all scales simultaneously, so that the prevailing cascade direction cannot be revealed.  
Both direct and inverse cascades can be well pronounced during the emergence. However, we cannot support the point of view regarding the dominant role of the inverse cascade (merging  of small features into the larger ones) in the formation of an AR. Although the coalescence of small magnetic elements into larger pores and sunspots is observed, the prevailed energy contribution to the AR comes from emergence of large-scale structures. 

According to turbulence theory \citep[see, \textit{e.g.},][]{Biskamp1993}, the energy input at some scale interval is followed by redistribution of energy along the spectrum in both directions: as toward smaller scales (direct cascade), so to larger scales (inverse cascade). Energy redistribution along the spectrum requires some finite time. This can be a reason why in our study a signature of inverse cascade is better visible when data accumulation is applied (see Figure~\ref{fig6}). At the same time, highly intermittent nature of the nonlinear transfer function $\partial E(k,t)/ \partial t$ (see  Figure~\ref{fig4}) does not allow us to go in data accumulation very deeply. We presume that the HMI cadence of 12 min is a suitable time interval to reveal the cascade direction. Our visual analysis of magnetograms (not shown) as well as high-resolution observations of \cite{Bernasconi2002} suggested that small magnetic elements flowing around during the AR emergence have lifetimes from tens of minutes to hours and a linear size of order of 1 Mm, so we can catch majority of them. 

To conclude, our results suggest that emergence of an AR is an intermittent rather than a monotonous process. Short-term intervals of energy increase are followed by intervals of energy decrease at different scales that can be explained by the turbulent nature of photospheric plasma interacting with emerging magnetic flux. A direct and inverse cascades operate simultaneously and, for the majority of time, they contribute to the energy gain at all scales.

\begin{acks}
We are grateful to the anonymous referee whose comments helped us to improve the paper. SDO is a mission for NASA’s Living With a Star(LWS) program. The SDO/HMI data were provided by the Joint Science Operation Center (JSOC). This study was supported by the Russian Science Foundation, Project 18-12-00131.
\end{acks}


\textbf{Disclosure of Potential Conflicts of Interest} The authors declare that they have no conflicts of interest.


\end{article} 


\begin{thebibliography}{}


\bibitem[\protect\citeauthoryear{Abramenko}{2005}]
{Abramenko2005}
Abramenko, V.I.: 2005, Relationship between Magnetic Power Spectrum and Flare Productivity in Solar Active Regions. \apj\ {\bf 629}, 1141.
\href{https://doi.org/10.1086/431732}{DOI} \href{http://adsabs.harvard.edu/abs/2005ApJ...629.1141A}{ADS}


\bibitem[\protect\citeauthoryear{Abramenko \emph{et al.}}{2001}]
{Abramenko2001}
Abramenko, V., Yurchyshyn, V., Wang, H., Goode, P.R.: 2001, Magnetic Power Spectra Derived from Ground and Space Measurements of the Solar Magnetic Fields. \solphys\ {\bf 201}, 225. \href{https://doi.org/10.1023/A:1017544723973}{DOI} \href{http://adsabs.harvard.edu/abs/2001SoPh..201..225A}{ADS}


\bibitem[\protect\citeauthoryear{Bernasconi \emph{et al.}}{2002}]
{Bernasconi2002}
Bernasconi, P.N., Rust, D.M., Georgoulis, M.K., Labonte, B.J.: 2002, Moving Dipolar Features in an Emerging Flux Region. \solphys\ {\bf 209}, 119. 
\href{https://doi.org/10.1023/A:1020943816174}{DOI} \href{http://adsabs.harvard.edu/abs/2002SoPh..209..119B}{ADS}


\bibitem[\protect\citeauthoryear{Biskamp}{1993}]
{Biskamp1993}
Biskamp, D.: 1993, Nonlinear Magnetohydrodynamics, {\it Cambridge Monographs on Plasma Physics, Cambridge [England]; New York, NY: Cambridge University Press, |c1993}, p. 378.
\href{http://adsabs.harvard.edu/abs/1993noma.book.....B}{ADS}


\bibitem[\protect\citeauthoryear{Bobra \emph{et al.}}{2014}]
{Bobra2014}
Bobra, M.G., Sun, X., Hoeksema, J.T., Turmon, M., Liu, Y., Hayashi, K., Barnes, G., Leka, K.D.: 2014, The Helioseismic and Magnetic Imager (HMI) Vector Magnetic Field Pipeline: SHARPs - Space-Weather HMI Active Region Patches. \solphys\ {\bf 289}, 3549. \href{https://doi.org/10.1007/s11207-014-0529-3}{DOI} \href{https://ui.adsabs.harvard.edu/#abs/2014SoPh..289.3549B/abstract}{ADS}


\bibitem[\protect\citeauthoryear{Borrero \emph{et al.}}{2011}]
{Borrero2011}
Borrero, J.M., Tomczyk, S., Kubo, M., Socas-Navarro, H., Schou, J., Couvidat, S., Bogart, R.: 2011, VFISV: Very Fast Inversion of the Stokes Vector for the Helioseismic and Magnetic Imager. \solphys\ {\bf 273}, 267. \href{https://doi.org/10.1007/s11207-010-9515-6}{DOI} \href{https://ui.adsabs.harvard.edu/#abs/2011SoPh..273..267B/abstract}{ADS}


\bibitem[\protect\citeauthoryear{Brants}{1985}]
{Brants1985}
Brants, J.J.: 1985, High-resolution spectroscopy of active regions. II Line-profile interpretation, applied to an emerging flux region. {\solphys}\ {\bf 95}, 15.
\href{https://doi.org/10.1007/BF00162633}{DOI} \href{http://adsabs.harvard.edu/abs/1985SoPh...95...15B}{ADS}


\bibitem[\protect\citeauthoryear{Brun \emph{et al.}}{2015}]
{Brun2015}
Brun, A.S., Browning, M.K., Dikpati, M., Hotta, H., Strugarek, A.: 2015, Recent Advances on Solar Global Magnetism and Variability. \ssr\ {\bf 196}, 101. 
\href{https://doi.org/10.1007/s11214-013-0028-0}{DOI} \href{http://adsabs.harvard.edu/abs/2015SSRv..196..101B}{ADS}


\bibitem[\protect\citeauthoryear{Centeno}{2012}]
{Centeno2012}
Centeno, R.: 2012, The Naked Emergence of Solar Active Regions Observed with SDO/HMI, \apj\ {\bf 759}, 72.
\href{https://doi.org/10.1088/0004-637X/759/1/72}{DOI} \href{http://adsabs.harvard.edu/abs/2012ApJ...759...72C}{ADS}



\bibitem[\protect\citeauthoryear{Charbonneau}{2010}]
{Charbonneau2010}
Charbonneau, P.: 2010, Dynamo Models of the Solar Cycle. {\it Liv. Rev. Solar Phys.} {\bf 7}, 3.
\href{https://doi.org/10.12942/lrsp-2010-3}{DOI} \href{http://adsabs.harvard.edu/abs/2010LRSP....7....3C}{ADS}



\bibitem[\protect\citeauthoryear{Charbonneau}{2014}]
{Charbonneau2014}
Charbonneau, P.: 2014, Solar Dynamo Theory. {\it Ann. Rev. Astron. Astrophys.} {\bf 52}, 251.
\href{https://doi.org/10.1146/annurev-astro-081913-040012}{DOI} \href{http://adsabs.harvard.edu/abs/2014ARA%26A..52..251C}{ADS}


\bibitem[\protect\citeauthoryear{Chen, Rempel, and Fan}{2017}]
{Chen2017}
Chen, F., Rempel, M., Fan, Y.: 2017, Emergence of Magnetic Flux Generated in a Solar Convective Dynamo. I. The Formation of Sunspots and Active Regions, and The Origin of Their Asymmetries. {\apj} {\bf 846}, 149.
\href{https://doi.org/10.3847/1538-4357/aa85a0}{DOI} \href{http://adsabs.harvard.edu/abs/2017ApJ...846..149C}{ADS}


\bibitem[\protect\citeauthoryear{Cheung and Isobe}{2014}]
{Cheung2014}
Cheung, M.C.M., Isobe, H.: 2014, Flux Emergence (Theory). {\it Liv. Rev. Solar Phys.} {\bf 11}, 3.
\href{https://doi.org/10.12942/lrsp-2014-3}{DOI} \href{http://adsabs.harvard.edu/abs/2014LRSP...11....3C}{ADS}


\bibitem[\protect\citeauthoryear{Cheung \emph{et al.}}{2017}]
{Cheung2017}
Cheung, M.C.M., van Driel-Gesztelyi, L., Mart{\'{\i}}nez Pillet, V., Thompson, M.J.: 2017, The Life Cycle of Active Region Magnetic Fields. {\ssr} {\bf 210}, 317. 
\href{https://doi.org/10.1007/s11214-016-0259-y}{DOI} \href{http://adsabs.harvard.edu/abs/2017SSRv..210..317C}{ADS}


\bibitem[\protect\citeauthoryear{Couvidat \emph{et al.}}{2012a}]
{Couvidat2012a}
Couvidat, S., Rajaguru, S.P., Wachter, R., Sankarasubramanian, K., Schou, J., Scherrer, P.H.: 2012a, Line-of-Sight Observables Algorithms for the Helioseismic and Magnetic Imager (HMI) Instrument Tested with Interferometric Bidimensional Spectrometer (IBIS) Observations. {\solphys} {\bf 278}, 217. \href{https://doi.org/10.1007/s11207-011-9927-y}{DOI} \href{https://ui.adsabs.harvard.edu/#abs/2012SoPh..278..217C/abstract}{ADS}


\bibitem[\protect\citeauthoryear{Couvidat \emph{et al.}}{2012b}]
{Couvidat2012b}
Couvidat, S., Schou, J., Shine, R.A., Bush, R.I., Miles, J.W., Scherrer, P.H., Rairden, R.L.: 2012b, Wavelength Dependence of the Helioseismic and Magnetic Imager (HMI) Instrument onboard the Solar Dynamics Observatory (SDO). {\solphys} {\bf 275}, 285. \href{https://doi.org/10.1007/s11207-011-9723-8}{DOI} \href{https://ui.adsabs.harvard.edu/#abs/2012SoPh..275..285C/abstract}{ADS}


\bibitem[\protect\citeauthoryear{Fan}{2009}]
{Fan2009}
Fan, Y.: 2009, Magnetic Fields in the Solar Convection Zone. {\it Liv. Rev. Solar Phys.} {\bf 6}, 4.
\href{https://doi.org/10.12942/lrsp-2009-4}{DOI} \href{http://adsabs.harvard.edu/abs/2009LRSP....6....4F}{ADS}


\bibitem[\protect\citeauthoryear{Fan and Fang}{2014}]
{Fan2014}
Fan, Y., Fang, F.: 2014, A Simulation of Convective Dynamo in the Solar Convective Envelope: Maintenance of the Solar-like Differential Rotation and Emerging Flux. {\apj} {\bf 789}, 35.
\href{https://doi.org/10.1088/0004-637X/789/1/35}{DOI} \href{http://adsabs.harvard.edu/abs/2014ApJ...789...35F}{ADS}


\bibitem[\protect\citeauthoryear{Hewett \emph{et al.}}{2008}]
{Hewett2008}
Hewett, R.J., Gallagher, P.T., McAteer, R.T.J., Young, C.A., Ireland, J., Conlon, P.A., Maguire, K.: 2008, Multiscale Analysis of Active Region Evolution. {\solphys} {\bf 248}, 311.
\href{https://doi.org/10.1007/s11207-007-9028-0}{DOI} \href{http://adsabs.harvard.edu/abs/2008SoPh..248..311H}{ADS}


\bibitem[\protect\citeauthoryear{Hoeksema \emph{et al.}}{2014}]
{Hoeksema2014}
Hoeksema, J.T., Liu, Y., Hayashi, K., Sun, X., Schou, J., Couvidat, S., Norton, A., Bobra, M., Centeno, R., Leka, K.D., Barnes, G., Turmon, M.: 2014, The Helioseismic and Magnetic Imager (HMI) Vector Magnetic Field Pipeline: Overview and Performance. {\solphys} {\bf 289}, 3483. \href{https://doi.org/10.1007/s11207-014-0516-8}{DOI} \href{https://ui.adsabs.harvard.edu/#abs/2014SoPh..289.3483H/abstract}{ADS}


\bibitem[\protect\citeauthoryear{Kutsenko and Abramenko}{2016}]
{Kutsenko2016}
Kutsenko, A.S., Abramenko, V.I.: 2016, Using SDO/HMI Magnetograms as a Source of the Solar Mean Magnetic Field Data. {\solphys} {\bf 291}, 1613. \href{https://doi.org/10.1007/s11207-016-0940-z}{DOI} \href{https://ui.adsabs.harvard.edu/#abs/2016SoPh..291.1613K/abstract}{ADS}


\bibitem[\protect\citeauthoryear{Leka \emph{et al.}}{2009}]
{Leka2009}
Leka, K.D., Barnes, G., Crouch, A.D., Metcalf, T.R., Gary, G.A., Jing, J., Liu, Y.: 2009, Resolving the 180$^{\circ}$ Ambiguity in Solar Vector Magnetic Field Data: Evaluating the Effects of Noise, Spatial Resolution, and Method Assumptions. {\solphys} {\bf 260}, 83. \href{https://doi.org/10.1007/s11207-009-9440-8}{DOI} \href{https://ui.adsabs.harvard.edu/#abs/2009SoPh..260...83L/abstract}{ADS}


\bibitem[\protect\citeauthoryear{Leka, Barnes, and Wagner}{2017}]
{Leka2017}
Leka, K.D., Barnes, G., Wagner, E.L.: 2017, Evaluating (and Improving) Estimates of the Solar Radial Magnetic Field Component from Line-of-Sight Magnetograms. {\solphys} {\bf 292}, 36. \href{https://doi.org/10.1007/s11207-017-1057-8}{DOI} \href{http://adsabs.harvard.edu/abs/2017SoPh..292...36L}{ADS}


\bibitem[\protect\citeauthoryear{Lites, Skumanich, and Martinez Pillet}{1998}]
{Lites1998}
Lites, B.W., Skumanich, A., Martinez Pillet, V.: 1998, Vector magnetic fields of emerging solar flux. I. Properties at the site of emergence. {\aap} {\bf 333}, 1053. 
\href{http://adsabs.harvard.edu/abs/1998A%26A...333.1053L}{ADS}



\bibitem[\protect\citeauthoryear{Liu et al.}{2012}]
{Liu2012}
Liu Y., Hoeksema J.T., Scherrer P.H., Schou J., Couvidat S., Bush R.I., Duvall T.L., Hayashi K., Sun X., Zhao X., 2012, Comparison of Line-of-Sight Magnetograms Taken by the Solar Dynamics Observatory/Helioseismic and Magnetic Imager and Solar and Heliospheric Observatory/Michelson Doppler Imager. {\solphys} {\bf 279}, 295.
\href{https://doi.org/10.1007/s11207-012-9976-x}{DOI}
\href{http://adsabs.harvard.edu/abs/2012SoPh..279..295L}{ADS}


\bibitem[\protect\citeauthoryear{Mandage and McAteer}{2016}]
{Mandage2016}
Mandage, R.S., McAteer, R.T.J.: 2016, On the Non-Kolmogorov Nature of Flare-productive Solar Active Regions. {\apj} {\bf 833}, 237. \href{https://doi.org/10.3847/1538-4357/833/2/237}{DOI}
\href{https://ui.adsabs.harvard.edu/#abs/2016ApJ...833..237M/abstract}{ADS}


\bibitem[\protect\citeauthoryear{Metcalf}{1994}]
{Metcalf1994}
Metcalf, T.R.: 1994, Resolving the 180-degree ambiguity in vector magnetic field measurements: The `minimum' energy solution. {\solphys} {\bf 155}, 235. \href{https://doi.org/10.1007/BF00680593}{DOI}
\href{https://ui.adsabs.harvard.edu/#abs/1994SoPh..155..235M/abstract}{ADS}


\bibitem[\protect\citeauthoryear{Norton \emph{et al.}}{2006}]
{Norton2006}
Norton, A.A., Graham, J.P., Ulrich, R.K., Schou, J., Tomczyk, S., Liu, Y., Lites, B.W., L{\'o}pez Ariste, A., Bush, R.I., Socas-Navarro, H., Scherrer, P.H.: 2006, Spectral Line Selection for HMI: A Comparison of Fe~{\sc i} 6173 \AA\ and Ni~{\sc i} 6768 \AA.  {\solphys} {\bf 239}, 69. \href{https://doi.org/10.1007/s11207-006-0279-y}{DOI}
\href{https://ui.adsabs.harvard.edu/#abs/2006SoPh..239...69N/abstract}{ADS}


\bibitem[\protect\citeauthoryear{Scherrer \emph{et al.}}{1995}]
{Scherrer1995}
Scherrer, P.H., Bogart, R.S., Bush, R.I., Hoeksema, J.T., Kosovichev, A.G., Schou, J., Rosenberg, W., Springer, L., Tarbell, T.D., Title, A., Wolfson, C.J., Zayer, I., MDI Engineering Team: 1995, The Solar Oscillations Investigation - Michelson Doppler Imager. {\solphys} {\bf 162}, 129. \href{https://doi.org/10.1007/BF00733429}{DOI} \href{https://ui.adsabs.harvard.edu/#abs/1995SoPh..162..129S/abstract}{ADS}


\bibitem[\protect\citeauthoryear{Scherrer \emph{et al.}}{2012}]
{Scherrer2012}
Scherrer, P.H., Schou, J., Bush, R.I., Kosovichev, A.G., Bogart, R.S., Hoeksema, J.T., Liu, Y., Duvall, T.L., Zhao, J., Title, A.M., Schrijver, C.J., Tarbell, T.D., Tomczyk, S.: 2012, The Helioseismic and Magnetic Imager (HMI) Investigation for the Solar Dynamics Observatory (SDO). {\solphys} {\bf 275}, 207.
\href{https://doi.org/10.1007/BF00733429}{DOI} \href{https://ui.adsabs.harvard.edu/#abs/1995SoPh..162..129S/abstract}{ADS}


\bibitem[\protect\citeauthoryear{Schou \emph{et al.}}{2012}]
{Schou2012}
Schou, J., Scherrer, P.H., Bush, R.I., Wachter, R., Couvidat, S., Rabello-Soares, M.C., Bogart, R.S., Hoeksema, J.T., Liu, Y., Duvall, T.L., Akin, D.J., Allard, B.A., Miles, J.W., Rairden, R., Shine, R.A., Tarbell, T.D., Title, A.M., Wolfson, C.J., Elmore, D.F., Norton, A.A., Tomczyk, S.: 2012, Design and Ground Calibration of the Helioseismic and Magnetic Imager (HMI) Instrument on the Solar Dynamics Observatory (SDO). {\solphys} {\bf 275}, 229. \href{https://doi.org/10.1007/s11207-011-9842-2}{DOI} \href{http://adsabs.harvard.edu/abs/2012SoPh..275..229S}{ADS}


\bibitem[\protect\citeauthoryear{Strous and Zwaan}{1999}]
{Strous1999}
Strous, L.H., Zwaan, C.: 1999, Phenomena in an Emerging Active Region. II. Properties of the Dynamic Small-Scale Structure. {\apj} {\bf 527}, 435.
\href{https://doi.org/10.1086/308071}{DOI} \href{http://adsabs.harvard.edu/abs/1999ApJ...527..435S}{ADS}


\bibitem[\protect\citeauthoryear{van Driel-Gesztelyi and Green}{2015}]
{vanDrielGesztelyi2015}
van Driel-Gesztelyi, L., Green, L.M.: 2015, Evolution of Active Regions. {\it Liv. Rev. Solar Phys.} {\bf 12}, 1.
\href{https://doi.org/10.1007/lrsp-2015-1}{DOI} \href{http://adsabs.harvard.edu/abs/2015LRSP...12....1V}{ADS}


\bibitem[\protect\citeauthoryear{Zwaan}{1985}]
{Zwaan1985}
Zwaan, C.: 1985, The emergence of magnetic flux. {\solphys} {\bf 100}, 397.
\href{https://doi.org/10.1007/BF00158438}{DOI} \href{http://adsabs.harvard.edu/abs/1985SoPh..100..397Z}{ADS}


\end{thebibliography}
\end{document}